\documentclass{eptcs}
\providecommand{\event}{Axel Legay (Ed.): International Workshop on Verification of Infinite-State Systems (INFINITY)} 
\providecommand{\volume}{13}   
\providecommand{\anno}{2009}   
\providecommand{\firstpage}{1} 
\providecommand{\eid}{1}       

\newtheorem{definition}{Definition}
\newtheorem{theorem}{Theorem}
\newtheorem{proposition}{Proposition}
\newtheorem{lemma}{Lemma}
\newenvironment{proof}{\textbf{Proof.}\ }{\hfill\fbox{}
\medskip}

\usepackage{amsmath, amssymb} 
\usepackage{url}
\usepackage{calc}

\usepackage{tikz}
\usetikzlibrary{arrows} 

\definecolor{gris}{rgb}{0.6,0.6,0.6}
\definecolor{grisfonce}{rgb}{0.2,0.2,0.2}
\definecolor{bleuciel}{rgb}{0.90,0.95,1}
\definecolor{vertclair}{rgb}{0.9, 1, 0.9}
\definecolor{vertfonce}{rgb}{0, 0.6, 0}
\definecolor{bleuclair}{rgb}{0.9, 0.9, 1}
\definecolor{bleufonce}{rgb}{0, 0, 0.6}
\definecolor{rougeclair}{rgb}{1, 0.9, 0.9}
\definecolor{rougefonce}{rgb}{0.6, 0, 0}
\definecolor{cpale1}{rgb}{1, 0.6, 0.6}
\definecolor{cpale2}{rgb}{0.6, 1, 0.6}\definecolor{cpale3}{rgb}{0.6, 0.6, 1}
\definecolor{cpale4}{rgb}{1, 1, 0.6}
\definecolor{cpale5}{rgb}{1, 0.6, 1}
\definecolor{cpale6}{rgb}{0.6, 1, 1}
\definecolor{cpale7}{rgb}{1, 0.4, 0.7}
\definecolor{cpale8}{rgb}{0.7, 0.4, 1}
\definecolor{cpale9}{rgb}{0.35, 1, 0.70}

\newcommand{\caml}{Caml}

\newcommand{\maxplus}{max--plus}

\newcommand{\absurde}{\emph{reductio ad absurdum}}

\newcommand{\mdpPolItParam}{\textit{P-mdpPI}}
\newcommand{\imperator}{\textsc{ImpRator}}
\newcommand{\mdpPolIt}{\textit{mdpPI}}
\newcommand{\mdpValueDet}{\textit{mdpVD}}
\newcommand{\mdpValueDetParam}{\textit{P-mdpVD}}

\newcommand{\maxPlusPolIt}{\textit{maxPI}}
\newcommand{\maxPlusPolicyImpr}{\textit{maxPImpr}}
\newcommand{\maxPlusParam}{\textit{P-maxPI}}
\newcommand{\maxPlusValueDet}{\textit{maxVD}}
\newcommand{\maxPlusValueDetParam}{\textit{P-maxVD}}

\newcommand{\Eta}{H}

\newcommand{\grandr}{{\mathbb R}}
\newcommand{\grandrmax}{{\mathbb R}_\mathit{max}}
\newcommand{\zero}{\epsilon}

\DeclareMathOperator*{\argmax}{arg\,max}

\newcommand{\we}{w}
\newcommand{\We}{W}
\newcommand{\va}{v}
\newcommand{\Va}{V}
\newcommand{\prob}{\mathit{Prob}}
\newcommand{\w}[2]{w_{#1,#2}}
\newcommand{\W}[2]{W_{#1,#2}}

\newenvironment{affectations}{\begin{tabular}{@{} l @{\ $:=$\ } l}}{\end{tabular}}

\newcommand{\ind}[1]{\noindent \hspace*{#1mm}\hspace*{#1mm}\hspace*{#1mm}\hspace*{#1mm}\hspace*{#1mm}\hspace*{#1mm}\hspace*{#1mm}\hspace*{#1mm}\hspace*{#1mm}\hspace*{#1mm}}
\newcommand{\vecteur}[1]{\left (
   \begin{array}{c}
      #1
   \end{array}
   \right )
}

\title{An Inverse Method for Policy-Iteration Based Algorithms\thanks{This work is partially supported by
the Agence Nationale de la Recherche, grant
ANR-06-ARFU-005, and by Institut Farman (ENS Cachan).}
}

\author{
\'Etienne Andr\'e
\institute{LSV, ENS Cachan \& CNRS \\ Cachan, France}
\email{andre@lsv.ens-cachan.fr}
\and
Laurent Fribourg
\institute{LSV, ENS Cachan \& CNRS \\ Cachan, France}
\email{fribourg@lsv.ens-cachan.fr}
}

\def\titlerunning{An Inverse Method for Policy-Iteration Based Algorithms}
\def\authorrunning{\'E. Andr\'e \& L. Fribourg}

\begin{document}
\maketitle

\begin{abstract}
We present an extension of two policy-iteration based algorithms on weighted graphs (viz., Markov Decision Problems and Max-Plus Algebras).
This extension allows us to solve the following {\em inverse problem}:
considering the weights of the graph to be unknown constants or \emph{parameters},
we suppose that a reference instantiation of those weights is given, and we aim at computing a constraint on the parameters
under which an optimal policy for the reference instantiation is still optimal.
The original algorithm is thus guaranteed to behave well around the reference instantiation, which provides us with some criteria of robustness.
We present an application of both methods to simple examples.
A prototype implementation has been done.
\end{abstract}

\section{Introduction}

We consider the {\em inverse problem} initially defined in the context of timed models.
More precisely, this inverse problem was first formalized in the context of Timing Constraint Graphs~\cite{ef08},
and then in the context of Timed Automata~\cite{ad94, acef09}.
We present here this problem in the context of systems modeled by
directed graphs with (parametric) weights associated to their edges, and more specifically in the cases of Markov Decision Processes (MDPs)~\cite{b57,howard60} and Max-Plus Algebras~\cite{ccggq98}.

Let us first present the {\em direct}
problem in this context.
The model is given under the form of a directed graph $G$,
with weights that are unknown constants or \emph{parameters}.
We also assume that a \emph{reference instantiation} $\pi_0$ is given for these parametric weights. Roughly speaking,
a {\em policy} is a function which associates with each state of the graph
an action which goes from the state to (a set of) successor state(s).
Each action has a specific weight.
The weight of a {\em path} (or sequence of actions) is the sum of the weights
of its constitutive actions.
The {\em value} (or cost) of a given policy $\mu$ for a given state $s$
corresponds to the mean weight of the paths induced by $\mu$, which go from $s$ to a final state of the graph.
Given a specific instantiation $\pi_0$ of the parameters,
the direct problem consists in computing
an {\em optimal} policy, that is a policy which gives the
{\em minimal value} (or \emph{maximal value}) when the parameters are instantiated with $\pi_0$.


The optimal policy is classically found using
the method of {\em policy iteration} ($\mathit{PI}$) (see \cite{howard60}).
The corresponding value is then computed by the {\em value
determination}
procedure ($\mathit{VD}$) (see, e.g., \cite{ccggq98}).
We show in this paper that the inverse problem can be simply stated, and solved via a natural generalization of the procedures of
policy iteration and value determination.
We  focus here on two classes of models:
Markov Decision Processes and Max-Plus Algebras.

Given a reference valuation $\pi_0$, the inverse algorithm generalizes the direct algorithm ``around'' $\pi_0$, and infers a constraint on the parameters guaranteeing a similar behavior as under $\pi_0$.
This ensures that the original algorithm continues to behave well around $\pi_0$, thus giving some criteria of \emph{robustness}.


We first give the general framework of our method (Sect.~\ref{s:framework}).
We then present the adaptation of the inverse method to
Markov Decision Processes (Sect.~\ref{s:mdp}) and Max-Plus Algebras (Sect.~\ref{s:maxplus}).
We conclude by giving some final remarks (Sect.~\ref{s:final}).

\section{General Framework}\label{s:framework}

\subsection{Preliminaries}\label{ss:general:prelim}

Throughout this paper, we assume a fixed set $P = \{p_1, \dots, p_{N} \}$  of \emph{parameters}.
A {\em parameter instantiation}~$\pi$ is a function
$\pi : P \rightarrow \grandr$ assigning a real constant to each parameter.
There is a one-to-one correspondence between instantiations and points in $\grandr^{N}$.
We will often identify an instantiation $\pi$ with the point $(\pi(p_1), \dots, \pi(p_{N}))$.

\begin{definition}\label{def:constraint}
 A \emph{linear inequality on the parameters $P$} is an inequality $e \prec e'$, where $\prec \in \{<, \leq\}$, and $e, e'$ are two \emph{linear terms} of the form
 $$ \Sigma_i \alpha_i p_i + d$$
 where $1 \leq i \leq N$, $\alpha_i \in \grandr $ and $d \in \grandr$.

 A \emph{(convex) constraint on the parameters $P$}
  is a conjunction of inequalities on $P$.
\end{definition}

We say that a parameter instantiation $\pi$ \emph{satisfies} a constraint $K$ on the parameters, denoted by $\pi \models K$,
if the expression obtained by replacing each parameter $p$ in $K$ with $\pi(p)$ evaluates to true.
We will consider $\mathit{True}$ as a constraint on the parameters,
corresponding to the set of all possible instances for $P$.

\subsection{Overview of the Inverse Method}\label{ss:general:problem}

We assume given a weighted graph,
and an algorithm $\mathit{PI}$ of policy iteration.
We define a \emph{parametric} version of the weighted graph, i.e., a weighted graph whose weights are unknown constants, or \emph{parameters}.
Given a parametric weighted graph $G$ and an instantiation $\pi$ of the parameters, we denote by $G[\pi]$ the (standard) weighted graph, where the parameters $p_i$ have been replaced by their instance $\pi(p_i)$.
For a given graph $G$, a given reference instance $\pi_0$ of the parameters,
and an optimal policy $\mu_0$ found by $\mathit{PI}$ for $G[\pi_0]$,
our goal is to generate a constraint $K_0$ on the parameters such that:
\begin{enumerate}
  \item $\pi_0 \models K_0$, and
  \item $\mu_0$ is optimal for $G[\pi]$,
for any instantiation $\pi$ satisfying $K_0$ (i.e., $\pi \models K_0$).
\end{enumerate}

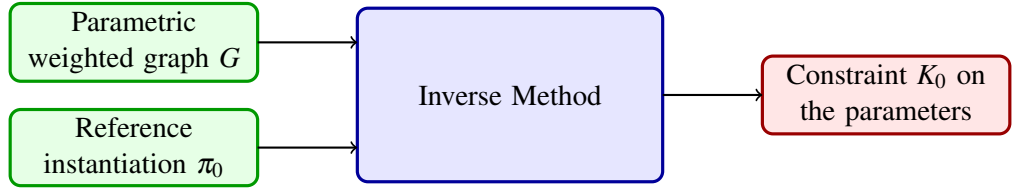
\begin{figure}
\tikzstyle{state}=[draw=gris, text=black, text width=8em, text centered, minimum height=2.5em, inner sep=3pt, outer sep=0pt, rounded corners, very thick]
\tikzstyle{input}=[state, fill=vertclair, draw=vertfonce] 
\tikzstyle{output}=[state, fill=rougeclair, draw=rougefonce]
\tikzstyle{imitator} = [state, text width=10em, fill=bleuclair, draw=bleufonce, minimum height=6em]
\def\blockdist{2.5}

{

\centering

\begin{tikzpicture} 
  \tikzstyle{every state}=[] 

    \node (imitator) at (5,1) [imitator] {Inverse Method};
    \node (output) at (10, 1) [output] {Constraint $K_0$ on the parameters};
    \node (input1) at (0,1.7) [state,input] {Parametric weighted graph $G$};
    \node (input2) at (0,0.3) [state,input] {Reference instantiation $\pi_0$};


    \path [draw, thick, ->] (input1) -- (imitator.west |- input1);
    \path [draw, thick, ->] (input2) -- (imitator.west |- input2);
    \path [draw, thick, ->] (imitator.east) -- (output);


\end{tikzpicture}

}

\caption{Our generic framework}
\label{fig:io}
\end{figure}

A trivial solution is $K_0 = \{ \pi_0 \}$.
However, our method will always generate something more general than $K_0 = \{ \pi_0 \}$, under the form of a conjunction of inequalities on the parameters (without any constant, apart from $0$).
Given $\mathit{PI}$, the framework of our inverse method is given in Fig.~\ref{fig:io}.
Given an algorithm $\mathit{PI}$ of policy iteration from the literature, calling itself an algorithm $\mathit{VD}$ of value determination,
our approach can be summarized as follows:
\begin{enumerate}
 \item Compute an optimal policy $\mu_0$ for the (standard) weighted graph $G[\pi_0]$, using $\mathit{PI}$;
 \item Compute a \emph{generic value} (or generic cost)
corresponding to $G$
for the policy $\mu_0$, using a parameterized version of $\mathit{VD}$;
 \item From the generic value computed above,
infer a constraint $K_0$ such that
$\mu_0$ is optimal for $G[\pi]$, for any instantiation
$\pi$ satisfying $K_0$.
\end{enumerate}

We now present such an inverse method
in the case of two policy-based iteration algorithms.

\section{Markov Decision Processes}\label{s:mdp}

\subsection{Preliminaries}\label{ss:mdp:prelim}

We consider in this section \emph{Markov Decision Processes}~\cite{b57} as an extension of weighted labeled directed graphs.
We associate to every edge of the graph a \emph{probability} such that, for a given state and a given action (or label), the sum of the probabilities of the edges leaving this state through this action is equal to~1.
Markov Decision Processes are widely used to model, e.g., the power consumption of devices (see, e.g., \cite{pbbm98}).
Formally:

\begin{definition}\label{def:mdp}
  A \emph{Markov Decision Process (MDP)} is a tuple
  $M = (S, A, \prob, \we)$, where
  \begin{itemize}
	\item $S = \{ s_1, \dots, s_n \}$ is a set of states;
	\item $A$ is a set of actions (or labels);
	\item $\prob : S \times A \times S \rightarrow [0, 1] $ is a probability function such that $\prob(s_1,a,s_2)$ is the probability that action $a$ in state $s_1$ will lead to state $s_2$,
		and $\forall s \in S, \forall a \in A : \Sigma_{s' \in S} \prob(s, a, s') = 1 $;
	\item $\we : S \times A \rightarrow \grandr$ is a weight function such that $\we(s,a)$ (also denoted by $\we_a(s)$) is the weight associated to the action $a$ when leaving~$s$.
  \end{itemize}
\end{definition}

In the following, we consider the MDP $M = (S, A, \prob, \we)$.
Given a state $s \in S$, we denote by $e(s)$ (for \emph{enabled}) the set of possible actions for $s$, i.e.,
$\{ a \in A \mid \exists s' \in S : \prob(s, a, s') > 0 \}$.
We suppose that, for any state $s \in S$, $e(s) \neq \emptyset$.
We also suppose that $M$ has a {\em unique} ``absorbing state'', i.e., a state which is reachable (with positive probability) from any other state for any policy, and which has a self-loop outgoing transition with weight 0 and probability 1.
We suppose in the following that the absorbing state is $s_n$.
For the sake of simplicity, we will not depict, in the graphs describing MDPs in this paper, the self-loop outgoing transition of the absorbing state.

In every state $s$ of $S \setminus \{ s_n \}$, we can choose \emph{non-deterministically} an action $a$ in $e(s)$.
Then, for this action, the system will evolve to a state $s'$ such that $\prob(s, a, s') > 0$.
A way of removing non-determinism from an MDP is to introduce a \emph{policy} $\mu$, i.e., a function from states to actions.
A policy is of the form $\mu = \{ s_1 \rightarrow a_{i_1}, s_2 \rightarrow a_{i_2}, \dots, s_{n-1} \rightarrow a_{i_{n-1}} \}$, with $a_{i_1}, \dots, a_{i_{n-1}} \in A$.
We denote by $\mu[s]$ the action associated to state~$s$.
The MDP, associated to a policy, behaves as a \emph{Markov chain}~\cite{kmst59}.

Given a policy $\mu$, the {\em associated value} is a function mapping each state $s$ to the mean sum of weights attached to the paths induced by $\mu$, which go from $s$ to $s_n$.
(By convention, the value associated to $s_n$ is null.)
A classical problem for MDPs is to find an \emph{optimal policy}, i.e., a policy under which the value function is maximum (or minimum), for  every
$s\in S$.
Note that, under the assumption of the existence of an absorbing state, such an optimal policy always exists, but is not necessarily unique (see, e.g., \cite{howard60}).
We focus here on finding an optimal policy for which the value function is \emph{minimal}.

We give in Fig.~\ref{algo:mdp:polIt} in Appendix~\ref{a:mdp} the classical algorithm \mdpPolIt{} for policy iteration on MDPs.
This algorithm computes the optimal policy for an MDP, and
it makes use of the algorithm \mdpValueDet{} for value determination in MDPs (see Fig.~\ref{algo:mdp:valueDet} in Appendix~\ref{a:mdp}).
Given an MDP and a policy, this second algorithm computes the mean sum of
weights attached to the paths reaching $s_n$,
for every starting state
in $S \setminus \{s_n\}$.
We denote by $\va[s]$ the value associated to state~$s$.
The value $\va$ computed by Algorithm \mdpValueDet{} is obtained by solving a system of linear equations, and is computed by applying
the inverse of a real-valued matrix to a parametric vector.
The fact that there is a single solution to this system is due to the fact that the matrix is invertible,
which comes itself from the existence of an absorbing state.

\subsection{An Illustrating Example} \label{ss:mdp:example}

Consider the case of a researcher getting by train from Paris to Bologna.
He can either take a night train \emph{Corail}, or use the French high-speed train TGV.
When there is no strike impacting the TGV service, the TGV usually needs 7 hours to go from Paris to Milan (with probability $4/5$).
It is then possible to take an Italian train, reaching Bologna from Milan in 1~hour with probability~1.
However, in case of strike (with probability~$1/5$), the TGV does not leave Paris, and the researcher should wait 7~more hours until the next TGV.
Note that this next TGV may also be on strike (with the same probability~$1/5$), and so on.
The night train can not be impacted by any strike, and it goes directly from Paris to Bologna in 11~hours with probability~1.

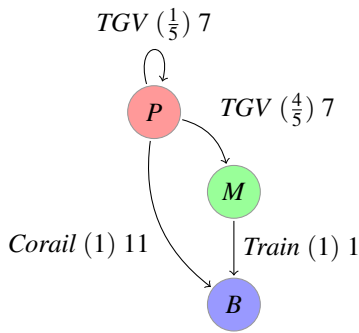
\begin{figure}
\centering
\small
\noindent
\begin{tikzpicture}[->, auto, node distance=1.5cm]
  \tikzstyle{state}=[circle, minimum size=20pt, inner sep=2pt, outer sep=1pt, draw=gris, text=black]

  \node[state,fill=cpale1] (p) {$P$};
  \node[state,fill=cpale2, below right of = p] (m) {$M$};
  \node[state,fill=cpale3, below of = m] (b) {$B$};

  \path (p)
		edge [loop above] node {$\mathit{TGV}$ $(\frac{1}{5})$ $7$} (p)
		edge [bend left] node {$\mathit{TGV}$ $(\frac{4}{5})$ $7$} (m)
		edge [below left, bend right] node {$\mathit{Corail}$ $(1)$ $11$} (b)
	(m)
 		edge node {$\mathit{Train}$ $(1)$ $1$} (b)
	;

\end{tikzpicture}
\caption{An example of Markov Decision Process}
\label{fig:ex-mdp}
\end{figure}

The MDP depicted in Fig.~\ref{fig:ex-mdp} summarizes those different possibilities, where $P$ stands for Paris, $M$ for Milan and $B$ for Bologna.
We denote by ``$\mathit{TGV}$ $(4/5)$ $7$'' a transition using label $\mathit{TGV}$ with probability $4/5$ and weight $7$ (i.e., 7~hours).
Note that the only source of non-determinism is in state $P$, where it is possible to choose between the \emph{TGV} and the \emph{Corail} actions.

We are first interested in the following question: considering the probability of strike, what is the best option, i.e., should we use the TGV or the night train\,?
This problem corresponds to finding an \emph{optimal policy} for this MDP, i.e., a policy minimizing the global weight of the system w.r.t. the probabilities.
An application of the (standard) algorithm \mdpPolIt{}~\cite{howard60} (see Fig.~\ref{algo:mdp:polIt} in Appendix~\ref{a:mdp}) to the MDP modeling the train journey from Paris to Bologna gives the following optimal policy:
$\mu = \{ P \rightarrow \emph{TGV}, M \rightarrow \emph{Train} \}$\footnote{
As $B$ is the absorbing state, recall that we do not define a policy for it.
}.
For this policy, the value for state $P$ (given by the last call to Algorithm \mdpValueDet{}), i.e., the expected time to reach Bologna, is $9.75$.

We now suppose that the train between Milan and Bologna can be subject to delays due, e.g., to some works on the track.
Our problem is the following:
until which delay of the train between Milan and Bologna the option ``TGV'' in Paris remains the best option?
In other words, until which delay of the train between Milan and Bologna the optimal policy remains optimal?
We are thus interested in computing a constraint on all the delays of the system, viewed as parameters, such that, for any instantiation of this constraint, the policy $\mu$ remains the optimal policy for this MDP.

\subsection{The Algorithm \mdpPolItParam{}}\label{ss:imperator}

We first adapt the notion of MDP to the parametric case.
We now consider that the weights of the MDP are \emph{parameters}.

\begin{definition}\label{def:pmdp}
  Given a set $P$ of parameters, a \emph{Parametric Markov Decision Process (PMDP)} is a tuple
  $M = (S, A, \prob, \We)$, where
  \begin{itemize}
	\item $S = \{ s_1, \dots, s_n \}$ is a set of states;
	\item $A$ is a set of actions;
	\item $\prob : S \times A \times S \rightarrow [0, 1] $ is a probability function such that $\prob(s_1,a,s_2)$ is the probability that action $a$ in state $s_1$ will lead to state $s_2$,
		and $\forall s \in S, \forall a \in A : \Sigma_{s' \in S} \prob(s, a, s') = 1 $;
	\item $\We : S \times A \rightarrow \mathit{P}$ is a parametric weight function such that $\We(s,a)$ (also denoted by $\We_a(s)$) is a parameter associated to the action $a$ when leaving~$s$.
  \end{itemize}
\end{definition}

We consider in the following the PMDP $M = (S, A, \prob, \We)$.
Given an instantiation $\pi$ of the parameters, we denote by $\We[\pi]$ the function from $S \times A$ to $\grandr$ obtained by replacing each occurrence of a parameter $p_i$ in $\We$ with the value $\pi(p_i)$, for $1 \leq i \leq N$.
By extension, we denote by $M[\pi]$ the (standard) MDP $(S, A, \prob, \We[\pi])$.

We first introduce the algorithm \mdpValueDetParam{}, given in Fig.~\ref{algo:mdp:valueDetParam}, which computes, given a policy $\mu$,
the parametric value associated to every state $s$ (i.e.,
the mean sum of the parametric weights of paths induced by $\mu$ going
from $s$ to $s_n$).
This algorithm is a straightforward adaptation to the parametric case of the classical algorithm \mdpValueDet{} of value determination for MDPs (see Fig.~\ref{algo:mdp:valueDet} in Appendix~\ref{a:mdp}).
We denote by $\Va[s]$ the parametric value associated to state~$s$.

\begin{figure}[!ht]
\centering
\fbox{
\begin{minipage}{0.9\textwidth}
\noindent
{\bf ALGORITHM \mdpValueDetParam{}$(M, \mu)$}

\smallskip

\noindent
\begin{tabular}{l @{\ \ \ } l @{\,:\ } l}
	\emph{Input} & $M$ & Parametric Markov Decision Process $(S, A, \prob, \We)$\\
		& $\mu$ & Policy \\
	\emph{Output} & $\Va$ & Parametric value function \\
\end{tabular}

\smallskip

{\bf SOLVE} $\{ \Va[s] = \We_{\mu[s]}(s) + \sum_{s' \in S} \prob(s, \mu[s], s') \times \Va[s'] \} _ {s \in S\setminus\{s_n\}}$

\end{minipage}
}

\caption{Algorithm for parametric value determination for MDPs}
\label{algo:mdp:valueDetParam}
\end{figure}

The value $\Va$ computed by this algorithm \mdpValueDetParam{} is obtained by solving a system of linear equations.
Since this system is of the form $\Va = A \times \Va + B$, it is equivalent to $\Va = (1 - A)^{-1} \times B$, and can be implemented using the inversion of matrix $(1 - A)$.
Note that this matrix $A$ is computed from matrix $\prob$ and vector $\mu$, and is therefore a constant real-valued matrix (i.e., containing no parameters).
As for the algorithm \mdpValueDet{}, the fact that there is a single solution to this system comes from the existence of an absorbing state.
Note also that the parametric value associated to a state is a \emph{linear term}, as defined in Def.~\ref{def:constraint}.

We state in the following Lemma that, given $M$ and $\mu$, the instantiation with $\pi$ of the parametric value associated to $M$ w.r.t. $\mu$ is equal to the value associated to $M[\pi]$ w.r.t. $\mu$.
We use $\Va[\pi]$ to denote the parametric value $\Va$ instantiated with $\pi$.

\begin{lemma}\label{lemma:mdp:valuedet}
  Let $M = (S, A, \prob, \We)$ be a PMDP,
  $\pi$ an instantiation of the parameter,
  and $\mu$ a policy for $M$.
  Let $\Va = \mdpValueDetParam(M, \mu)$.
  Then $\Va[\pi] = \mdpValueDet(M[\pi], \mu)$.
\end{lemma}

\begin{proof}
  The algorithm $\mdpValueDetParam(M, \mu)$ consists in solving a system of the form $\Va = A \times \Va + \We_{\mu[s]}$.
  Hence, $\Va = (1 - A) ^ {-1} \times \We_{\mu[s]}$.
  Moreover, the algorithm $\mdpValueDet(M[\pi], \mu)$ consists in solving a system of the form $\va = A' \times \va + \We[\pi]_{\mu[s]}$, i.e.,  $\va = (1 - A') ^ {-1} \times \We[\pi]_{\mu[s]}$.
  It is easy to see on the two algorithms that $A = A'$.
  We trivially have: for all $s$, $\We[\pi]_{\mu[s]}(s) = (\We_{\mu[s]}(s))[\pi]$, where $(\We_{\mu[s]}(s))[\pi]$ denotes the linear term $\We_{\mu[s]}(s)$ where every occurrence of a parameter $p_i$ was replaced by its instantiation $\pi_i$.
  Hence, $\Va[\pi] = \mdpValueDet(M[\pi], \mu)$.
\end{proof}

We now introduce the algorithm \mdpPolItParam{}, which fits in our general framework of Fig.~\ref{fig:io}.
Given a reference instantiation $\pi_0$ of the parameters, this algorithm takes as input a PMDP $M$, and an optimal policy $\mu_0$ associated to $M[\pi_0]$ (which can be computed using $\mdpPolIt(M[\pi_0])$).
Recall that, by ``optimal'', we mean here a policy under which the value of states is {\em minimal}.
The algorithm outputs a constraint $K_0$ on the parameters such that:
\begin{enumerate}
 \item $\pi_0 \models K_0$, and
 \item for any $\pi \models K_0$, $\mu_0$ is an optimal policy of $M[\pi]$. 
\end{enumerate}



\begin{figure} 
\centering
\fbox{
\begin{minipage}{0.9\textwidth}
\noindent
{\bf ALGORITHM \mdpPolItParam{}($M, \mu_0$)}

\smallskip

\noindent
\begin{tabular}{l @{\ \ \ } l @{\,:\ } l}
	\emph{Input} & $M$ & Parametric Markov Decision Process $(S, A, \prob, \We)$\\
		& $\mu_0$ & Optimal policy for the reference instantiation of the parameters \\
	\emph{Output} & $K_0$ & Constraint on the set of parameters \\
\end{tabular}

\medskip

\ind{0}
\begin{affectations}
  $\Va $ & $ $\mdpValueDetParam$(M, \mu_0)$\\
  $K_0 $ & $ \mathit{True}$\\
\end{affectations}

\smallskip

\ind{0}	\textbf{FOR EACH} $s \in S\setminus\{s_n\}$ \textbf{DO}

	\ind{1} \textbf{FOR EACH} $a \in e(s)$ s.t. $a \neq \mu_0[s]$ \textbf{DO}


		\ind{2} $K_0 := K_0 \land \{ \We_a(s) + \sum_{s' \in S} \prob(s, a, s') \Va[s'] \geq \Va[s] \}$

%

	\ind{1}	\textbf{OD}\\
\ind{0}	\textbf{OD}

\end{minipage}
}

\caption{Algorithm solving the inverse problem for MDPs}
\label{algo:imperator}
\end{figure}

The algorithm \mdpPolItParam{} is given in Fig.~\ref{algo:imperator}.
We can summarize this algorithm as follows:
\begin{enumerate}
 \item Compute the parametric value function, which associates to any state a parametric value w.r.t. $\mu_0$, using Algorithm \mdpValueDetParam{};
 \item For every state $s\neq s_n$, for every action $a$ different from the action $\mu_0[s]$ given by the the optimal policy, generate the following inequality stating that $a$ is not a better action (i.e.,
an action which would lead to a better policy) than $\mu_0[s]$:
  $$ \We_a(s) + \sum_{s' \in S}\prob(s, a, s') \Va[s'] \geq \Va[s]$$
\end{enumerate}
The above set of inequalities implies that, for any $s$ and $a$,
the policy obtained from $\mu_0$ by changing $\mu_0[s]$ with $a$,
does not improve policy $\mu_0$
(i.e., does not lead to any smaller value of state).

\subsection{Properties}\label{ss:mdp:properties}

We first show that $\pi_0$ models the constraint $K_0$ output by our algorithm.

\begin{proposition}\label{prop:mdp:pi0-k0}
Let  $\mu_0 =\mdpPolIt{}(M[\pi_0])$, and
 $K_0 = $\mdpPolItParam{}$(M, \mu_0)$.
  Then $\pi_0 \models K_0$.
\end{proposition}

\begin{proof}(By \absurde{})
  Suppose $\pi_0 \not\models K_0$.
  Then, there exists an inequality $J$ in $K_0$ such that $\pi_0 \not\models J$.
  By construction, this inequality $J$ is of the form
  $ \We_a(s) + \sum_{s' \in S}\prob(s, a, s') \Va[s']  \geq \Va[s]$, for some $s$ and some $a$.
  If this inequality $J$ is not satisfied by $\pi_0$, this means that $a$ is a strictly better policy for $s$ than the policy $\mu_0[s]$ in $M[\pi_0]$, which is not possible since $\mu_0$ is an optimal policy for $M[\pi_0]$.
\end{proof}

\begin{proposition}\label{prop:mdp:termination}
  Algorithm \mdpPolItParam{} terminates.
\end{proposition}

\begin{proof}
  Since $M$ contains exactly one absorbing state, the computation of the parametric value in \mdpValueDetParam{} 
  is guaranteed to terminate with a single solution.
  Since the number of generated inequalities is finite, it is easy to see that Algorithm \mdpPolItParam{} terminates.
\end{proof}

%
%

Note that  the size (in term of number of inequalities) of the constraint $K_0$ output by our algorithm is in $O(|S| \times |A|)$,
where $|S|$ (resp. $|A|$) denotes the number of states (resp. actions) of $M$.

\smallskip


We now state that our algorithm \mdpPolItParam{} solves the inverse problem as described in Sect.~\ref{ss:general:problem}.

\begin{theorem}\label{th:mdp}
Let  $\mu_0 =\mdpPolIt{}(M[\pi_0])$, and
  $K_0 = \mdpPolItParam{}(M, \mu_0)$.
  Then:
  \begin{enumerate}
   \item $\pi_0 \models K_0$, and
   \item for all $\pi \models K_0$, policy $\mu_0$ is optimal for $M[\pi]$.
  \end{enumerate}

\end{theorem}

\begin{proof} 
  Let us prove item (2) by \absurde{}.
  Recall that $M = (S, A, \prob, \We)$.
  Let 
  $\pi \models K_0$.
  We have $M[\pi] = (S, A, \prob, \We[\pi])$.

  Suppose that $\mu_0$ is not an optimal policy for $M[\pi]$.
  Let $\mu$ be an optimal policy for $M[\pi]$.
  Then there exists some state $s$ such that $\mu[s]$ is a strictly better policy than $\mu_0[s]$ for $M[\pi]$.
  Let $a = \mu[s]$ and $a_0 = \mu_0[s]$.
  Let $\va = \mdpValueDet(M[\pi], \mu)$.
  Since $a$ is a strictly better policy than $a_0$ for state $s$ in $M[\pi]$, then, from the last iteration of Algorithm $\mdpPolIt(M[\pi])$, we have:
	$\We[\pi]_{a_0}(s) + \sum_{s' \in S} \prob(s, a_0, s') \va[s']  > \We[\pi]_a(s) + \sum_{s' \in S} \prob(s, a, s') \va[s']$.

  Moreover, since $a \neq \mu_0[s]$, Algorithm $\mdpPolItParam(M, \mu_0)$ generates the following inequality in $K_0$:
	$ \We_a(s) + \sum_{s' \in S} \prob(s, a, s') \Va[s'] \geq \Va[s] $.
  Since $\Va[s] =  \We_{a_0}(s) + \sum_{s' \in S} \prob(s, a_0, s') \times  \Va[s'] $ (from the call to Algorithm $\mdpValueDetParam(M, \mu_0)$),
  this inequality is equal to
	$ \We_a(s) + \sum_{s' \in S} \prob(s, a, s') \Va[s'] \geq \We_{a_0}(s) + \sum_{s' \in S} \prob(s, a_0, s') \times  \Va[s'] $.
  Since $\pi \models K_0$, the instantiation of $K_0$, and in particular of this inequality,  with~$\pi$ should evaluate to true.
  By Lemma~\ref{lemma:mdp:valuedet}, we have $\Va[\pi] = \va$.
  Hence, by instantiating the inequality with~$\pi$, we get:
	$ \We[\pi]_a(s) + \sum_{s' \in S} \prob(s, a, s') \va[s'] \geq \We[\pi]_{a_0}(s) + \sum_{s' \in S} \prob(s, a_0, s') \times  \va[s'] $,
  which is exactly the contrary of what was stated before.

\end{proof}

\subsection{Application to the Example}

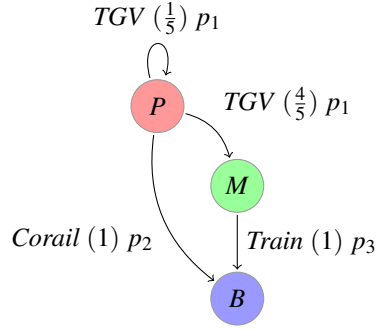
\begin{figure}
\centering
\small
\noindent
\begin{tikzpicture}[->, auto, node distance=1.5cm]
  \tikzstyle{state}=[circle, minimum size=20pt, inner sep=2pt, outer sep=1pt, draw=gris, text=black]

  \node[state,fill=cpale1] (p) {$P$};
  \node[state,fill=cpale2, below right of = p] (m) {$M$};
  \node[state,fill=cpale3, below of = m] (b) {$B$};

  \path (p)
		edge [loop above] node {$\mathit{TGV}$ $(\frac{1}{5})$ $p_1$} (p)
		edge [bend left] node {$\mathit{TGV}$ $(\frac{4}{5})$ $p_1$} (m)
		edge [below left, bend right] node {$\mathit{Corail}$ $(1)$ $p_2$} (b)
	(m)
 		edge node {$\mathit{Train}$ $(1)$ $p_3$} (b)
	;

\end{tikzpicture}
\caption{An example of Parametric Markov Decision Process}
\label{fig:ex-pmdp}
\end{figure}

Consider again the journey from Paris to Bologna described in Sect.~\ref{ss:mdp:example}.
We give in Fig.~\ref{fig:ex-pmdp} the PMDP $M$ adapted from Fig.~\ref{fig:ex-mdp} to the parametric case.
The set of parameters is $P = \{ p_1, p_2, p_3 \}$.
The reference instantiation $\pi_0$ of the parameters is the following one\footnote{
From the definition of the MDP and PMDP, the weight corresponding to leaving state $P$ through action $\mathit{TGV}$ must be the same for any destination state.
This is the reason why, in state $P$, the duration corresponding to waiting the next train (7~hours) is the same as the time needed to reach Milan.
In the case where we would need different weights, it is possible to set an average value for the weight by taking into account the respective probabilities.
}:

\smallskip

{\centering

\begin{tabular}{r @{\ = \ } r @{\ \ \ \ \ \ \ \ } r @{\ = \ } r @{\ \ \ \ \ \ \ \ } r @{\ = \ } r }
$p_1$ & 7 &
$p_2$ & 11 &
$p_3$ & 1
\end{tabular}

}

\smallskip

Note that $M[\pi_0]$ corresponds to the (standard) MDP depicted in Fig.~\ref{fig:ex-mdp}.

Let us briefly explain the application of \mdpPolItParam{} to this example.
We first compute the optimal policy $\mu_0$ for $M[\pi_0]$.
As said in Sect.~\ref{ss:mdp:example}, $\mu_0 = \{ P \rightarrow \emph{TGV}, M \rightarrow \emph{Train} \}$.
Applying Algorithm $\mdpValueDetParam(M, \mu_0)$, we then compute the parametric value of each state w.r.t. the optimal policy $\mu_0$.
As $B$ is an absorbing state, we have $\Va[B] = 0$.
Thus, we trivially have $\Va[M] = p_3$.
We then have $\Va[P] = \We_{\mu_0[P]}(P) + 1/5 \times  \Va[P] + 4/5 \times  \Va[M]$,
which gives $\Va[P] = 5/4 \times  p_1 + p_3$.
Note that, by replacing the parameters $p_i$ by $\pi_0(p_i)$ in $\Va[P]$ for $i = 1, 2, 3$, we get
$5/4 \times  7 + 1 = 9.75$, which is equal to the value computed by the classical algorithm \mdpPolIt{} (from Lemma~\ref{lemma:mdp:valuedet}).

We now compute the constraint $K_0$.
The only non-determinism being in state~$P$, we generate the following inequality:
$1 \times  (p_2 + \Va[B]) \geq \Va[P]$, which gives:
$$ p_2 \geq \frac{5}{4} p_1 + p_3 $$
By instantiating all the parameters except the one corresponding to the duration of the train between Milan and Bologna (i.e., $p_3$),
we get the following inequality:
$$ p_3 \leq \frac{9}{4} $$
Thus, if the train between Milan and Bologna takes more than 2h15 (i.e., is impacted by a delay of more that 1h15), then the optimal policy of the TGV will not be optimal anymore, and we should consider another option.

\paragraph{Remark.}
This example being simple, it was rather easy to predict this result from the direct application of the classical algorithm \mdpPolIt{} to the MDP described in Fig.~\ref{fig:ex-mdp}.
Indeed, the expected value $\va[P]$ in state~$P$ is equal to $9.75$ so, if a delay of more than $11 - 9.75 = 1.25$ (i.e., 1h15) occurs somewhere between Paris and Bologna using the TGV option (in particular between Milan and Bologna), the TGV policy will not be optimal anymore.
Our algorithm \mdpPolItParam{} is of course interesting for more complex systems.

\subsection{Implementation}\label{ss:tool}

The algorithm \mdpPolItParam{} has been implemented under the form of a program named \imperator{}
(standing for \emph{Inverse Method for Policy with Reward AbstracT behaviOR}).
This program, containing about 4300~lines of code,
is written in \caml{},
and uses matrix inversion to compute the parametric value $\Va$ in Algorithm \mdpValueDetParam{}.
We applied our program to various examples of MDPs modeling devices.
For a system containing 11~states, 4 actions and 132~transitions, corresponding to the model of a robot evolving in a bounded physical space~\cite{sb03}, our program \imperator{} generates a constraint in~$0.17$\,s.

The program and various case studies can be downloaded on the \imperator{} Web page\footnote{\url{http://www.lsv.ens-cachan.fr/~andre/ImPrator/}}.

\section{Max--Plus Algebra}\label{s:maxplus}

We consider in this section the algorithm 4.4 ``Max--Plus Policy Iteration'' of~\cite{ccggq98} (which will be here denoted by \maxPlusPolIt{}), used to compute the maximal circuit mean of a weighted directed graph in the framework of \maxplus{} algebra.
We are interested in computing a constraint on the weights attached to a directed graph, such that the circuit of maximal mean remains the same, under any instantiation satisfying this constraint.

We use in this section a formalism similar to the one in~\cite{ccggq98}\footnote{However, we will denote the policy by $\mu$ instead of $\pi$, both in order to keep the formalism introduced previously and in order to avoid confusion with $\pi$, standing in our framework for an instantiation of the parameters.}.

\subsection{Preliminaries}\label{ss:prelim}

The max--plus semiring $\grandrmax$ is the set $\grandr \cup \{ - \infty \}$,
equipped with $\emph{max}$ and ${+}$.
The zero element will be denoted by $\zero$ ($\zero = - \infty$).
The unit element will not be used in this paper.

\begin{definition}
A \emph{directed weighted graph} (or \emph{DWG}) $G$ is a triple $(S, E, w)$, where:
\begin{itemize}
 \item $S$ is a finite set of states,
 \item $E$ is a set of oriented edges $E \subset S \times S$,
 \item $w : E \rightarrow \grandr$ is a function associating to every edge a real-valued weight.
\end{itemize}
We denote by $w(e)$, or alternatively by $\w{i}{j}$, the weight associated to the edge $e = (i, j)$.
We associate to $G$ a \emph{matrix} $M \in (\grandrmax)^{S \times S}$, such that
$$ M_{ij} = \left \{ \begin{array}{l l}
	\w{i}{j} & \text{if } (i, j) \in E,\\
	\zero & \text{otherwise} \\
\end{array}
 \right . $$
Conversely, we associate to any matrix $M \in \grandr^{n \times n}$ the graph $G_M = (S, E, w)$,
where $S = \{1, \dots, n\}$, $E = \{ (i, j) \in S \times S \mid M_{ij} \neq \zero \}$,
and $\w{i}{j} = M_{ij}$ for any $(i, j) \in S \times S$.
\end{definition}

In the following, we will mainly consider the formalism of matrices rather than the graphs.
We consider in the following the matrix $M$, whose associated graph $G_M$ is strongly connected.

\begin{definition}
Given a \emph{DWG} $G = (S, E, w)$, the \emph{maximal circuit mean} is
$$ \rho = \max_{c} \frac{\sum_{e \in c} w(e) }{ \sum _ {e \in c} 1} \ , $$
where the max is taken over all the circuits $c$ of $G$, and the sums are taken over all the edges $e$ of $c$.
\end{definition}

Note that, in the definition of $\rho$, the numerator is the weight of $c$, and the denominator is the length of $c$.

In the context of DWGs, given a matrix $M$, a \emph{policy} is a function $\mu$ from $S$ to $E$, such that for all~$i \in S$, $\mu[i]$ is an edge starting from $i$.
In the following, without loss of understanding, we will sometimes abbreviate the edge $\mu[i] = (i, j)$ as its target state $j$.
Given a policy $\mu$ for $M$, we denote by $\mu[i]$ the policy associated to state~$i$.
Moreover, we denote by $M^\mu$ the matrix such that, for any $i,j$, $M^\mu_{ij} = M_{ij}$ if $j = \mu[i]$, and $M^\mu_{ij} = \zero$ otherwise.

Given a matrix $M$ and a policy $\mu$, the \emph{value function}, denoted by $(\eta, x)$, associates to each state $i$ of~$S$ a couple $(\eta_i, x_i) \in \grandr \times \grandr$ (called ``(generalized) eigenmode'' in~\cite{ccggq98}).

An \emph{optimal} policy $\mu$ for $M$ induces a \emph{circuit} $c$ of maximal mean in graph $G$.
More precisely, $\mu[i]$ is an edge of $c$ if $i$ belongs to $c$, and there is a path from $i$ to a state of $c$ otherwise.
Moreover, the associated value $(\eta, x)$ is such that all the $\eta_i$s are identical, and equal to the \emph{maximal circuit mean}~$\rho$ of~$G$.\footnote{Note that the $\eta_i$s are also equal to the (unique) eigenvalue of $M$, and $x$ is an eigenvector of $M$ (see Theorem 3.1 in~\cite{ccggq98}).} 

The algorithm \maxPlusPolIt{} (see Fig.~\ref{algo:maxplus:polIt} in Appendix~\ref{a:maxplus}) computes an optimal policy for a given DWG.
Starting from an arbitrary policy, it iteratively improves the current policy using Algorithm \maxPlusPolicyImpr{} (see Fig.~\ref{algo:maxplus:policyImpr} in Appendix~\ref{a:maxplus}) 
and Algorithm \maxPlusValueDet{} (see Fig.~\ref{algo:maxplus:valueDet} in Appendix~\ref{a:maxplus}), which computes the associated value function $(\eta, x)$.

\subsection{An Illustrating Example}\label{ss:maxplus:example}

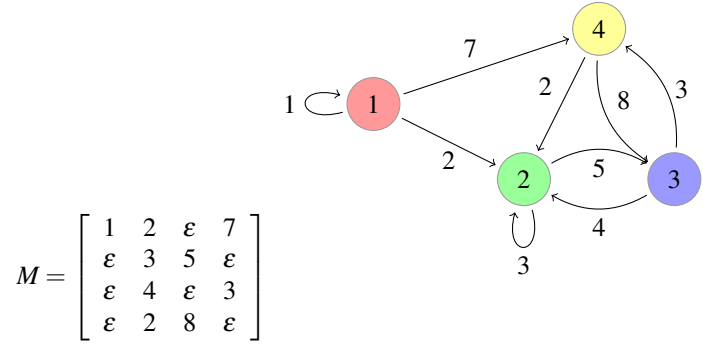
\begin{figure}
\centering

\small

$ M = \left [
   \begin{array}{c c c c}
      1 & 2 & \zero & 7 \\
      \zero & 3 & 5 & \zero \\
      \zero & 4 & \zero & 3 \\
      \zero & 2 & 8 & \zero \\
   \end{array}
\right ]$
\begin{tikzpicture}[->, auto]
  \tikzstyle{state}=[circle, minimum size=20pt, inner sep=1pt, outer sep=2pt, draw=gris, text=black]

  \node[state,fill=cpale1] (1) at (0, 1) {1};
  \node[state,fill=cpale2] (2) at (2, 0) {2};
  \node[state,fill=cpale3] (3) at (4, 0) {3};
  \node[state,fill=cpale4] (4) at (3, 2) {4};

  \path (1)
		edge [loop left] node {1} (1)
		edge [below] node {2} (2)
		edge node {7} (4)
	(2)
		edge [loop below] node {3} (2)
		edge [below, bend left] node {5} (3)
	(3)
		edge [bend left] node {4} (2)
		edge [right, bend right] node {3} (4)
	(4)
		edge [above left] node {2} (2)
		edge [bend right] node {8} (3)
	;
\end{tikzpicture}
\caption{A matrix and its graph}
\label{fig:ccggq98}
\end{figure}

We give in Fig.~\ref{fig:ccggq98} an example of DWG (coming from~\cite{ccggq98}) with its corresponding matrix.
We are interested in finding the maximal circuit mean of a DWG.
Let us briefly apply Algorithm \maxPlusPolIt{} to the matrix~$M$ of Fig.~\ref{fig:ccggq98}.
As in~\cite{ccggq98}, we choose the initial policy $\pi_1 : 1 \rightarrow 1, i \rightarrow 2$, for $i = 2, 3, 4$.
Applying Algorithm \maxPlusValueDet{}, we find a first circuit $c_1 : 1 \rightarrow 1$, with $\overline{\eta} = w(c_1) = 1$.
We set $\eta^1_1 = 1$, and $x^1_1 = 0$.
Since 1 is the only state which has access to 1, we apply algorithm \maxPlusValueDet{} to the subgraph of $G_{M}$ with states 2, 3, 4.
We find the circuit $c_2 : 2 \rightarrow 2$ and set $\overline{\eta} = w(c_2) = 3$, $\eta^1_2 = 3$, and $x^1_2 = 0$.
Since 3, 4 have access to 2, we set $\eta^1_i = 3$ for $i = 3, 4$.
Moreover, an application of (\ref{maxplus:eq:21}) yields $x^1_3 = 4 - 3 + x^1_2$, and $x^1_4 = 2 - 3 + x^1_2$.
To summarize:
$$
   \eta^1 = \vecteur{1 \\ 3 \\ 3 \\ 3 \\ }
   \ , \ \ \ \
   x^1 = \vecteur{0 \\ 0 \\ 0 \\ -1 \\ }
$$
We improve the policy using Algorithm \maxPlusPolicyImpr{}.
Since $J = \{ 1 \} \neq \emptyset$, we have a type 3a improvement.
This yields $\pi_2 : i \rightarrow 2$ for $i = 2, 3, 4$.
Only the entry 1 of $x^1$ and $\eta^1$ has to be modified, which yields

$$
   \eta^2 = \vecteur{3 \\ 3 \\ 3 \\ 3 \\ }
   \ , \ \ \ \
   x^2 = \vecteur{-1 \\ 0 \\ 1 \\ -1 \\ }
$$
We tabulate with less details the end of the run of the algorithm.
Algorithm \maxPlusPolicyImpr{}, type 3b, policy improvement.
$\pi_3 : 1 \rightarrow 4, 2 \rightarrow 3, 3 \rightarrow 2, 4 \rightarrow 3$.
Algorithm \maxPlusValueDet{}.
Circuit found $c : 3 \rightarrow  2 \rightarrow 3$,
$\overline{\eta} = (\w{2}{3} + \w{3}{2}) / 2 = 9/2$.
$$
   \eta^3 = \vecteur{\frac{9}{2} \\ \frac{9}{2} \\ \frac{9}{2} \\ \frac{9}{2} \\ }
   \ , \ \ \ \
   x^3 = \vecteur{\frac{11}{2} \\ 0 \\ -\frac{1}{2} \\ 3 \\ }
$$
Algorithm \maxPlusPolicyImpr{}, type 3b, policy improvement.
The only change is $\pi_4(3) = 4$.
Algorithm \maxPlusValueDet{}.
Circuit found $c : 3 \rightarrow 4 \rightarrow 3$,
$\overline{\eta} = (\w{3}{4} + \w{4}{3}) / 2 = 11/2$.
$$
   \eta^4 = \vecteur{\frac{11}{2} \\ \frac{11}{2} \\ \frac{11}{2} \\ \frac{11}{2} \\ }
   \ , \ \ \ \
   x^4 = \vecteur{4 \\ - \frac{1}{2} \\ 0 \\ \frac{5}{2} \\ }
$$
Algorithm \maxPlusPolicyImpr{}. Stop. 
Hence, we get the following result:

\begin{eqnarray}\label{eq:eigenmode}
   \eta = \vecteur{\frac{11}{2} \\ \frac{11}{2} \\ \frac{11}{2} \\ \frac{11}{2} \\ }
   \ , \ \ \ \ \
   x = \vecteur{4 \\ - \frac{1}{2} \\ 0 \\ \frac{5}{2} \\ }
   \ , \ \ \ \ \
   \mu = \vecteur{4 \\ 3 \\ 4 \\ 3 \\ }
\end{eqnarray}

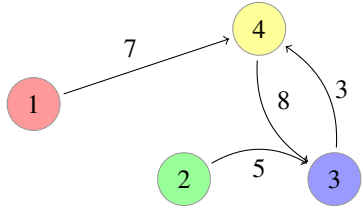
\begin{figure}
\centering

\small

\begin{tikzpicture}[->, auto]
  \tikzstyle{state}=[circle, minimum size=20pt, inner sep=1pt, outer sep=2pt, draw=gris, text=black]

  \node[state,fill=cpale1] (1) at (0, 1) {1};
  \node[state,fill=cpale2] (2) at (2, 0) {2};
  \node[state,fill=cpale3] (3) at (4, 0) {3};
  \node[state,fill=cpale4] (4) at (3, 2) {4};

  \path (1)
		edge node {7} (4)
	(2)
		edge [below, bend left] node {5} (3)
	(3)
		edge [right, bend right] node {3} (4)
	(4)
		edge [bend right] node {8} (3)
	;
\end{tikzpicture}
\caption{The graph corresponding to the matrix $M^\mu$}
\label{fig:ccggq98-4}
\end{figure}

Thus, $11/2$ is an eigenvalue of $M$, and $x$ is an eigenvector.
The subgraph $M^\mu$ of $M$ restricted to the policy $\mu$ is given in Fig.~\ref{fig:ccggq98-4}.
We note that the mean of circuit $4 \rightarrow 3 \rightarrow 4$ is $(8 + 3) / 2 = 11/2$,
and it is easy to check that this circuit has the maximal circuit mean of the graph associated to $M$.

We are now interested in the following problem.
Suppose that one wants to minimize the weight associated to the edge $4 \rightarrow 3$ (of weight $\w{4}{3} = 8$).
What is the minimal value for $\w{4}{3}$ so that circuit $4 \rightarrow 3 \rightarrow 4$ remains the circuit of maximal mean in the graph~$M$\,?
In other words, we are interested in computing a constraint on the weights of the system, viewed as \emph{parameters}, so that the circuit of maximal mean remains the circuit of maximal mean.

\subsection{The Algorithm \maxPlusParam{}}\label{ss:maxplus:algo}

We first adapt the notion of DWG to the parametric case.
We now consider that the weights of the graph are \emph{parameters}.

\begin{definition} \label{def:pdwg}
Given a set $P$ of parameters, a \emph{parametric directed weighted graph} (or \emph{PDWG}) $G$ is a triple $(S, E, w)$, where:
\begin{itemize}
 \item $S$ is a finite set of states,
 \item $E$ is a set of oriented edges $E \subset S \times S$,
 \item $W : E \rightarrow P$ is a parametric function associating to every edge a parametric weight.
\end{itemize}
We denote by $W(e)$, or alternatively by $\W{i}{j}$, the parametric weight associated to the edge $e = (i, j)$.
We associate to $G$ a \emph{parametric matrix} $M \in (P \cup \zero)^{S \times S}$, such that
$$ M_{ij} = \left \{ \begin{array}{l l}
	\W{i}{j} & \text{if } (i, j) \in E,\\
	\zero & \text{otherwise} \\
\end{array}
 \right . $$
Conversely, we associate to any parametric matrix $M \in (P \cup \zero)^{n \times n}$ the graph $G_M = (S, E, W)$,
where $S = \{1, \dots, n\}$, $E = \{ (i, j) \in S \times S \mid M_{ij} \neq \zero \}$,
and $\W{i}{j} = M_{ij}$ for any $(i, j) \in S \times S$.
\end{definition}

We consider in the following the PDWG $(S, E, W)$, and its associated matrix $M$.
Given an instantiation $\pi$ of the parameters, we denote by $W[\pi]$ the weight function from $E$ to $\grandr$ obtained by replacing each occurrence of a parameter $p_i$ in $W$ with the value $\pi(p_i)$, for $1 \leq i \leq N$.
Similarly, we denote by $M[\pi]$ the matrix (in $(\grandrmax)^{n \times n}$) obtained by replacing in $M$ each occurrence of a parameter $p_i$ with the value $\pi(p_i)$, for $1 \leq i \leq N$.
The notion of policy can be extended to the parametric framework in a natural way.

\begin{figure}
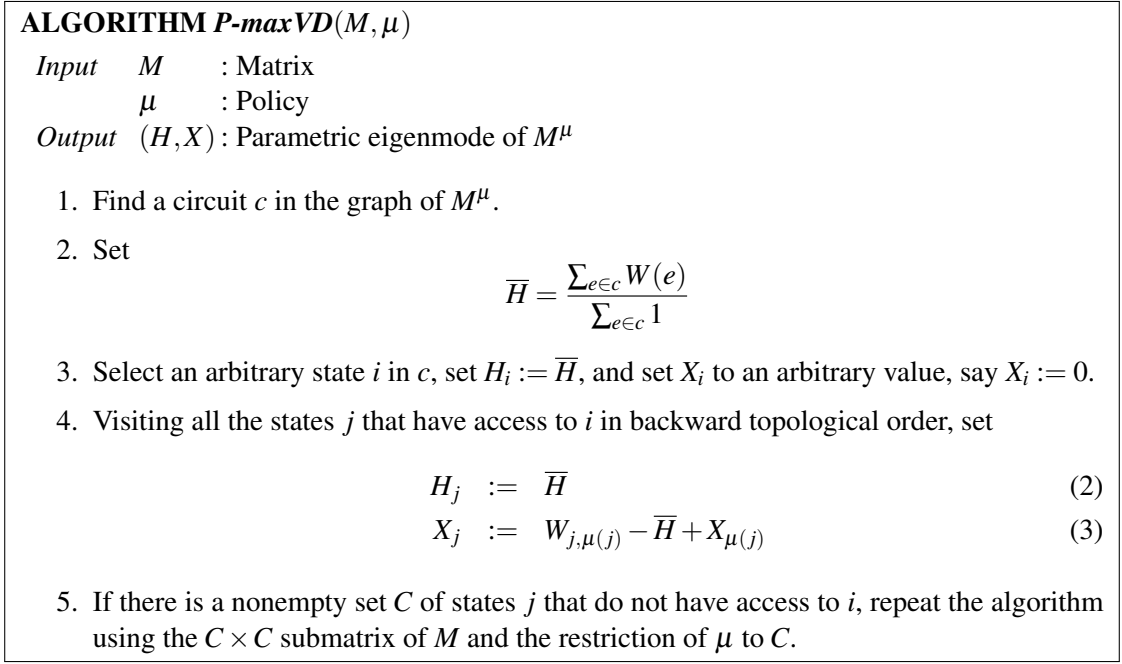

\centering
\fbox{
\begin{minipage}{0.9\textwidth}
\noindent
{\bf ALGORITHM $\maxPlusValueDetParam(M, \mu)$}

\smallskip

\noindent
\begin{tabular}{l @{\ \ \ } l @{\,:\ } l}
	\emph{Input} & $M$ & Matrix \\
		& $\mu$ & Policy \\
	\emph{Output} & $(\Eta, X)$ & Parametric eigenmode of $M^\mu$ \\
\end{tabular}

\smallskip

\begin{enumerate}
 \item Find a circuit $c$ in the graph of $M^\mu$.

 \item Set
$$\overline{H} = \frac {\sum_{e \in c} W(e)} {\sum_{e \in c} 1} $$

\item Select an arbitrary state $i$ in $c$,
set $\Eta_i := \overline{\Eta}$,
and set $X_i$ to an arbitrary value, say $X_i := 0$.

\item Visiting all the states $j$ that have access to $i$ in backward topological order, set
\begin{eqnarray}
\Eta_j & := & \overline{\Eta}\\
X_j & := & \W{j}{\mu(j)} - \overline{\Eta} + X_{\mu(j)} \label{eq:21}
\end{eqnarray}

\item If there is a nonempty set $C$ of states $j$ that do not have access to $i$,
repeat the algorithm using the $C \times C$ submatrix of $M$
and the restriction of $\mu$ to $C$.
\end{enumerate}

\end{minipage}
}
\caption{Algorithm for parametric value determination for maximal circuit mean}
\label{algo:maxplus:valueDetParam}
\end{figure}

Following the idea of our framework of Sect.~\ref{s:framework},
we first give in Fig.~\ref{algo:maxplus:valueDetParam} the algorithm \maxPlusValueDetParam{}.
This algorithm is an adaptation to the parametric case of the algorithm for value determination \maxPlusValueDet{} from~\cite{ccggq98} (see Fig.~\ref{algo:maxplus:valueDet} in Appendix~\ref{a:maxplus}).
Given a policy $\mu$, it computes a \emph{parametric} eigenmode $(\Eta, X)$ of $M^\mu$.
In other words, it associates to every state $i$ of $M^\mu$ two parametric values $\Eta_i$ and $X_i$, which are two linear terms (as defined in Def.~\ref{def:constraint}).

\begin{figure}
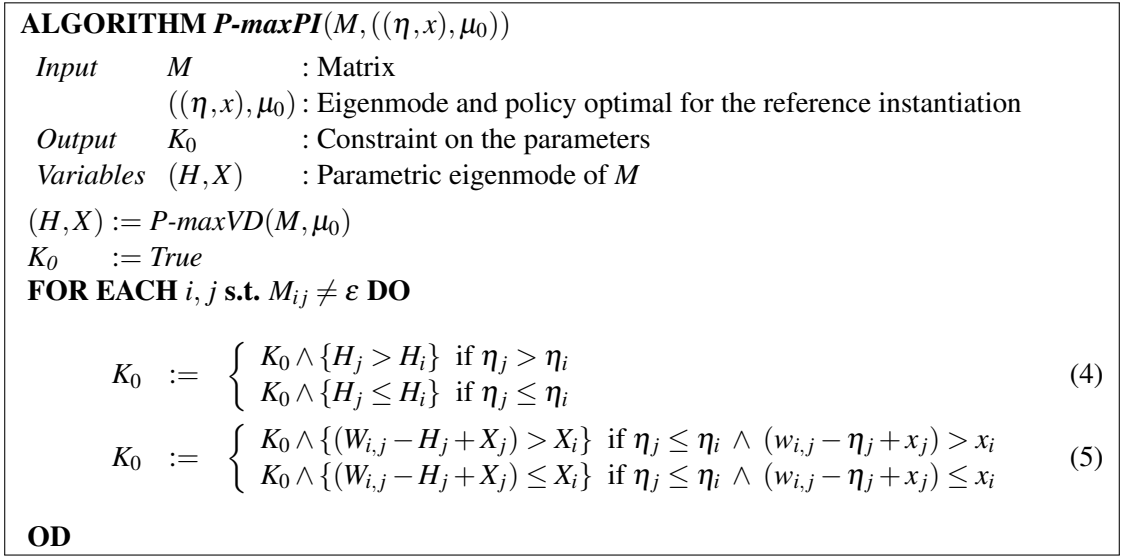

\centering
\fbox{
\begin{minipage}{0.9\textwidth}
\noindent
{\bf ALGORITHM \maxPlusParam$(M, ((\eta, x), \mu_0))$}

\smallskip

\noindent
\begin{tabular}{l @{\ \ \ } l @{\,:\ } l}
	\emph{Input} 	& $M$ & Matrix \\
			& $((\eta, x), \mu_0)$ & Eigenmode and policy optimal for the reference instantiation \\ 
	\emph{Output}	& $K_0$ & Constraint on the parameters \\
	\emph{Variables}
			& $(\Eta, X)$ & Parametric eigenmode of $M$\\
\end{tabular}

\smallskip

\ind{0} \begin{affectations}
	$(\Eta, X)$ & \maxPlusValueDetParam$(M, \mu_0)$ \\
	$\mathit{K_0}$ & $\mathit{True}$ \\
        \end{affectations}

\ind{0} \textbf{FOR EACH} $i, j$ \textbf{s.t.} $M_{ij} \neq \zero$ \textbf{DO}
	\begin{eqnarray}
	 K_0 & := & \label{ineq:1}
	\left \{
	\begin{array}{l @{\ \ \text{if}\ } l}
	K_0 \land \{ \Eta_j > \Eta_i \} & \eta_j > \eta_i \\
	K_0 \land \{ \Eta_j \leq \Eta_i \} & \eta_j \leq \eta_i \\
	\end{array} \right.
	\\
	 K_0 & := & \label{ineq:2}
	\left \{
	\begin{array}{l @{\ \ \text{if}\ } l}
	K_0 \land \{ (\W{i}{j} - \Eta_j + X_j) > X_i \} & \eta_j \leq \eta_i \ \land \ (\w{i}{j} - \eta_j + x_j) > x_i \\
	K_0 \land \{ (\W{i}{j} - \Eta_j + X_j) \leq X_i \} & \eta_j \leq \eta_i \ \land \  (\w{i}{j} - \eta_j + x_j) \leq x_i \\
	\end{array} \right.
	\end{eqnarray}
\ind{0} \textbf{OD}

\end{minipage}
}

\caption{Algorithm solving the maximal circuit mean inverse problem}
\label{algo:maxplus:param}
\end{figure}

We now introduce the algorithm \maxPlusParam{}, which fits in our general framework of Fig.~\ref{fig:io}.
We give the algorithm \maxPlusParam{} in Fig.~\ref{algo:maxplus:param}.
As in the MDP Section, we first apply the standard algorithm for policy iteration from the literature, i.e., we first call Algorithm \maxPlusPolIt{}, given in Fig.~\ref{algo:maxplus:polIt} in Appendix~\ref{a:maxplus} (which makes itself use of Algorithms \maxPlusValueDet{} and \maxPlusPolicyImpr{}, available in Fig.~\ref{algo:maxplus:valueDet} and Fig.~\ref{algo:maxplus:policyImpr} respectively).
This algorithm computes the eigenmode $(\eta, x)$ of the maximal circuit mean of $M$, and the corresponding policy $\mu_0$.
Then we compute the parametric eigenmode of $M$ associated to $\mu_0$, using Algorithm \maxPlusValueDetParam{}.
Finally, we compute a set of inequalities ensuring that the policy $\mu_0$ is the optimal policy w.r.t. maximal circuit mean.
This generation of inequalities is the adaptation to the parametric case of the test of optimality performed in the classical algorithm \maxPlusPolicyImpr{} (given in Fig.~\ref{algo:maxplus:policyImpr} in Appendix~\ref{a:maxplus}).

We now state that our algorithm \maxPlusParam{} solves the inverse problem as described in Sect.~\ref{ss:general:problem}.

\begin{theorem}\label{th:maxplus}
  Let $((\eta, x), \mu_0)=\maxPlusPolIt{}(M[\pi_0])$ and
$K_0 = $\maxPlusParam{}$(M, ((\eta, x), \mu_0))$.
  Then:
  \begin{enumerate}
   \item $\pi_0 \models K_0$, and
   \item for all $\pi \models K_0$, policy $\mu_0$ corresponds to a maximal mean circuit of $M[\pi]$.
  \end{enumerate}

\end{theorem}


Note that, although we guarantee that the circuit of maximal mean in $M[\pi]$ is always the same, for any $\pi \models K$, the mean value itself varies with $\pi$. 

\subsection{Application to the Example}

Let us apply the algorithm \maxPlusParam{} given in Fig.~\ref{algo:maxplus:param} to the graph from~\cite{ccggq98} depicted in Fig.~\ref{fig:ccggq98} in Sect.~\ref{ss:maxplus:example}.

We first apply Algorithm \maxPlusPolIt{}, which gives the result (\ref{eq:eigenmode}) of Sect.~\ref{ss:maxplus:example}.

Then, we call Algorithm \maxPlusValueDetParam{}.
The circuit $c$ found is $3 \rightarrow 4 \rightarrow 3$.
We set $\overline{\Eta} = (\W{3}{4} + \W{4}{3}) / 2$.
We then pick up, say, state $3$ in $c$, set $\overline{\Eta_3} := \overline{\Eta}$ and $X_3 := 0$.
Then, visiting all the states $j$ that have access to $i$ in backward topological order, we have:
\begin{itemize}
 \item For state $1$: $\Eta_1 = \overline{\Eta}$, and $X_1 = \W{1}{4} - \overline{\Eta} + X_4$
 \item For state $2$: $\Eta_2 = \overline{\Eta}$, and $X_2 = \W{2}{3} - \overline{\Eta} + X_3$
 \item For state $4$: $\Eta_4 = \overline{\Eta}$, and $X_4 = \W{4}{3} - \overline{\Eta} + X_3$
\end{itemize}
Since the set $C$ of states $j$ that do not have access to $i$ is empty, the algorithm \maxPlusValueDetParam{} terminates.
After resolution of the system above, we get
$$
   \Eta = \vecteur{\frac{\W{3}{4} + \W{4}{3}}{2} \\
		\frac{\W{3}{4} + \W{4}{3}}{2} \\
		\frac{\W{3}{4} + \W{4}{3}}{2} \\
		\frac{\W{3}{4} + \W{4}{3}}{2} \\
	}
   \ , \ \ \ \
   X = \vecteur{\W{1}{4} - \W{3}{4} \\
	 \W{2}{3} - \frac{1}{2}\W{3}{4} - \frac{1}{2}\W{4}{3} \\
	 0 \\
	 \frac{1}{2}\W{4}{3} - \frac{1}{2}\W{3}{4} \\ }
$$

We now generate the inequalities.
For every edge $(i, j)$ of the graph, we generate two inequalities, i.e., inequalities $(\ref{ineq:1})$ and $(\ref{ineq:2})$ of the algorithm \maxPlusParam{}.
All generated inequalities, including the trivial ones (i.e., of the form $a \leq a$, for some linear term $a$), are depicted in Fig.~\ref{fig:generation}.
For every edge $(i, j)$ of the graph, we first give the inequality corresponding to $(\ref{ineq:1})$, and then the inequality corresponding to $(\ref{ineq:2})$.
The conjunction of those inequalities gives the constraint $K_0$ output by the algorithm.

\begin{figure}
{

\centering

\footnotesize

\begin{tabular}{| c | r @{\ } c @{\ } l |}

	\hline
      $1 \rightarrow 1$ & $ \frac{1}{2} \W{3}{4} + \frac{1}{2} \W{4}{3} $ & $\leq$ & $ \frac{1}{2} \W{3}{4} + \frac{1}{2} \W{4}{3}$ \\
      	& $ \W{1}{1} - \frac{1}{2} \W{3}{4} - \frac{1}{2} \W{4}{3} + \W{1}{4} - \frac{1}{2} \W{3}{4} - \frac{1}{2} \W{4}{3} + \W{4}{3} - \frac{1}{2} \W{3}{4} - \frac{1}{2} \W{4}{3} $ & $\leq$ & $ \W{1}{4} - \frac{1}{2} \W{3}{4} - \frac{1}{2} \W{4}{3} + \W{4}{3} - \frac{1}{2} \W{3}{4} - \frac{1}{2} \W{4}{3}$ \\

	\hline

      $1 \rightarrow 2$ & $ \frac{1}{2} \W{3}{4} + \frac{1}{2} \W{4}{3} $ & $\leq$ & $ \frac{1}{2} \W{3}{4} + \frac{1}{2} \W{4}{3}$ \\
      	& $ \W{1}{2} - \frac{1}{2} \W{3}{4} - \frac{1}{2} \W{4}{3} + \W{2}{3} - \frac{1}{2} \W{3}{4} - \frac{1}{2} \W{4}{3} $ & $\leq$ & $ \W{1}{4} - \frac{1}{2} \W{3}{4} - \frac{1}{2} \W{4}{3} + \W{4}{3} - \frac{1}{2} \W{3}{4} - \frac{1}{2} \W{4}{3}$ \\

	\hline
      $1 \rightarrow 4$ & $ \frac{1}{2} \W{3}{4} + \frac{1}{2} \W{4}{3} $ & $\leq$ & $ \frac{1}{2} \W{3}{4} + \frac{1}{2} \W{4}{3}$ \\
      	& $ \W{1}{4} - \frac{1}{2} \W{3}{4} - \frac{1}{2} \W{4}{3} + \W{4}{3} - \frac{1}{2} \W{3}{4} - \frac{1}{2} \W{4}{3} $ & $\leq$ & $ \W{1}{4} - \frac{1}{2} \W{3}{4} - \frac{1}{2} \W{4}{3} + \W{4}{3} - \frac{1}{2} \W{3}{4} - \frac{1}{2} \W{4}{3}$ \\

	\hline
      $2 \rightarrow 2$ & $ \frac{1}{2} \W{3}{4} + \frac{1}{2} \W{4}{3} $ & $\leq$ & $ \frac{1}{2} \W{3}{4} + \frac{1}{2} \W{4}{3}$ \\
      	& $ \W{2}{2} - \frac{1}{2} \W{3}{4} - \frac{1}{2} \W{4}{3} + \W{2}{3} - \frac{1}{2} \W{3}{4} - \frac{1}{2} \W{4}{3} $ & $\leq$ & $ \W{2}{3} - \frac{1}{2} \W{3}{4} - \frac{1}{2} \W{4}{3}$ \\

	\hline
      $2 \rightarrow 3$ & $ \frac{1}{2} \W{3}{4} + \frac{1}{2} \W{4}{3} $ & $\leq$ & $ \frac{1}{2} \W{3}{4} + \frac{1}{2} \W{4}{3}$ \\
      	& $ \W{2}{3} - \frac{1}{2} \W{3}{4} - \frac{1}{2} \W{4}{3} + 0 $ & $\leq$ & $ \W{2}{3} - \frac{1}{2} \W{3}{4} - \frac{1}{2} \W{4}{3}$ \\

	\hline

      $3 \rightarrow 2$ & $ \frac{1}{2} \W{3}{4} + \frac{1}{2} \W{4}{3} $ & $\leq$ & $ \frac{1}{2} \W{3}{4} + \frac{1}{2} \W{4}{3}$ \\
      	& $ \W{3}{2} - \frac{1}{2} \W{3}{4} - \frac{1}{2} \W{4}{3} + \W{2}{3} - \frac{1}{2} \W{3}{4} - \frac{1}{2} \W{4}{3} $ & $\leq$ & $ 0$ \\

	\hline

      $3 \rightarrow 4$ & $ \frac{1}{2} \W{3}{4} + \frac{1}{2} \W{4}{3} $ & $\leq$ & $ \frac{1}{2} \W{3}{4} + \frac{1}{2} \W{4}{3}$ \\
      	& $ \W{3}{4} - \frac{1}{2} \W{3}{4} - \frac{1}{2} \W{4}{3} + \W{4}{3} - \frac{1}{2} \W{3}{4} - \frac{1}{2} \W{4}{3} $ & $\leq$ & $ 0$ \\

	\hline

      $4 \rightarrow 2$ & $ \frac{1}{2} \W{3}{4} + \frac{1}{2} \W{4}{3} $ & $\leq$ & $ \frac{1}{2} \W{3}{4} + \frac{1}{2} \W{4}{3}$ \\
      	& $ \W{4}{2} - \frac{1}{2} \W{3}{4} - \frac{1}{2} \W{4}{3} + \W{2}{3} - \frac{1}{2} \W{3}{4} - \frac{1}{2} \W{4}{3} $ & $\leq$ & $ \W{4}{3} -
\frac{1}{2} \W{3}{4} - \frac{1}{2} \W{4}{3}$ \\

	\hline

      $4 \rightarrow 3$ & $ \frac{1}{2} \W{3}{4} + \frac{1}{2} \W{4}{3} $ & $\leq$ & $ \frac{1}{2} \W{3}{4} + \frac{1}{2} \W{4}{3}$ \\
      	& $ \W{4}{3} - \frac{1}{2} \W{3}{4} - \frac{1}{2} \W{4}{3} + 0 $ & $\leq$ & $ \W{4}{3} - \frac{1}{2} \W{3}{4} - \frac{1}{2} \W{4}{3}$ \\

	\hline
\end{tabular}

}

\caption{The generation of $K_0$ for our example of graph}
\label{fig:generation}
\end{figure}

After simplification (trivially done by hand)
of the constraint $K_0$, we get the following constraint:

\smallskip

{

\centering
\small

\begin{tabular}{ c r @{\ } c @{\ } l }
  	  & $ 2\W{1}{1} $ & $\leq$ & $ \W{3}{4} + \W{4}{3} $ \\
  $\land$ & $ \W{1}{2} + \W{2}{3} $ & $\leq$ & $ \W{1}{4} + \W{4}{3} $ \\
  $\land$ & $ 2\W{2}{2} $ & $\leq$ & $ \W{3}{4} + \W{4}{3} $ \\
  $\land$ & $ \W{2}{3} + \W{3}{2} $ & $\leq$ & $ \W{3}{4} + \W{4}{3} $ \\
  $\land$ & $ 2\W{2}{3} + 2\W{4}{2} $ & $\leq$ & $ \W{3}{4} + 3\W{4}{3}$ \\
\end{tabular}

}

\smallskip

Recall that we were interested in Sect.~\ref{ss:maxplus:example} in knowing until which value it was possible to minimize $\W{4}{3}$ so that the circuit $4 \rightarrow 3 \rightarrow 4$ remained the circuit of maximal mean in the graph~$M$ of Fig.~\ref{fig:ccggq98}.
Let us instantiate all parameters except $\W{4}{3}$ in the constraint output by \maxPlusParam{}$(M, ((\eta, x), \mu_0))$.
We then get the following inequality:
$$ \W{4}{3} \geq 6 $$
Thus, provided this inequality is verified, the circuit $4 \rightarrow 3 \rightarrow 4$ remains the maximal mean circuit in the graph~$M$ of Fig.~\ref{fig:ccggq98}.
Note that it is actually easy to see on the graph in Fig.~\ref{fig:ccggq98} that, if $ \W{4}{3} < 6 $, the maximal mean circuit then becomes $2 \rightarrow 3 \rightarrow 2$, with $\eta = 9/2$.

\section{Final Remarks}\label{s:final}

We have presented an extension of two algorithms based
on policy-iteration for two models: Markov Decision Problems and Max-Plus Algebras.
For these models, we introduced a natural generalization of the policy-iteration method that solves the inverse problem, i.e:
considering the weights of the models to be unknown constants or \emph{parameters}, and given a reference instantiation $\pi_0$ of those weight parameters, we compute a constraint under which an optimal policy for $\pi_0$ is still optimal.
This increases our confidence in the robustness of policy-iteration based methods.

This inverse method was also experienced in another kind of weighted graphs, i.e., directed weighted graphs: in this context, we generate a constraint on the weights seen as parameters, guaranteeing that the shortest path from one state to another one remains the shortest path~\cite{and09b}.

Such an extension seems to work on several other policy-iteration algorithms.
In particular, we are studying the adaptation of the method to Markov decision processes with {\em two} weights, as used in the problem of \emph{dynamic power management}~\cite{pbbm98} for real-time systems where one wants to minimize the power consumption while keeping a certain level of efficiency.
We also plan to adapt the method to an extension of Algorithm \maxPlusPolIt{} allowing to treat deterministic games with
mean payoff~\cite{dg06}.

\paragraph{Acknowledgments.}
We thank an anonymous referee for his/her helpful comments.

\bibliographystyle{eptcs}
\bibliography{biblio}

\begin{thebibliography}{10}
\providecommand{\bibitemstart}[1]{\bibitem{#1}}
\providecommand{\bibitemend}{}
\providecommand{\bibliographystart}{}
\providecommand{\bibliographyend}{}
\providecommand{\url}[1]{\texttt{#1}}
\providecommand{\urlprefix}{Available at }
\providecommand{\bibinfo}[2]{#2}
\bibliographystart

\bibitemstart{ad94}
\bibinfo{author}{R.~Alur} \& \bibinfo{author}{D.~L. Dill}
  (\bibinfo{year}{1994}): \emph{\bibinfo{title}{A theory of timed automata}}.
\newblock {\sl \bibinfo{journal}{{TCS}}}
  \bibinfo{volume}{126}(\bibinfo{number}{2}), pp. \bibinfo{pages}{183--235}.
\bibitemend

\bibitemstart{and09b}
\bibinfo{author}{\'E. Andr\'e} (\bibinfo{year}{2009}):
  \emph{\bibinfo{title}{Une m\'ethode inverse pour les plus courts chemins}}.
\newblock \bibinfo{note}{Submitted to ETR '09}.
\bibitemend

\bibitemstart{acef09}
\bibinfo{author}{{\'E}tienne Andr{\'e}}, \bibinfo{author}{{\relax Th}omas
  Chatain}, \bibinfo{author}{Emmanuelle Encrenaz} \& \bibinfo{author}{Laurent
  Fribourg} (\bibinfo{year}{2009}): \emph{\bibinfo{title}{An Inverse Method for
  Parametric Timed Automata}}.
\newblock {\sl \bibinfo{journal}{International Journal of Foundations of
  Computer Science}}
  \urlprefix\url{http://www.lsv.ens-cachan.fr/Publis/PAPERS/PDF/ACEF-ijfcs09.p%
df}.
\newblock \bibinfo{note}{To appear}.
\bibitemend

\bibitemstart{b57}
\bibinfo{author}{R.~Bellman} (\bibinfo{year}{1957}): \emph{\bibinfo{title}{{A
  Markov decision process}}}.
\newblock {\sl \bibinfo{journal}{Journal of Mathematical Mechanics}}
  \bibinfo{volume}{6}, pp. \bibinfo{pages}{679--684}.
\bibitemend

\bibitemstart{ccggq98}
\bibinfo{author}{J.~Cochet-terrasson}, \bibinfo{author}{G.~Cohen},
  \bibinfo{author}{S.~Gaubert}, \bibinfo{author}{M.~Mc Gettrick} \&
  \bibinfo{author}{J-P. Quadrat} (\bibinfo{year}{1998}):
  \emph{\bibinfo{title}{Numerical Computation of Spectral Elements in Max-Plus
  Algebra}}.
\newblock In: {\sl \bibinfo{booktitle}{IFAC Conf. on Syst. Structure and
  Control}}.
\bibitemend

\bibitemstart{dg06}
\bibinfo{author}{V.~Dhingra} \& \bibinfo{author}{S.~Gaubert}
  (\bibinfo{year}{2006}): \emph{\bibinfo{title}{How to solve large scale
  deterministic games with mean payoff by policy iteration}}.
\newblock In: {\sl \bibinfo{booktitle}{Valuetools '06}}.
  \bibinfo{publisher}{ACM}, \bibinfo{address}{New York, NY, USA},
  p.~\bibinfo{pages}{12}.
\bibitemend

\bibitemstart{ef08}
\bibinfo{author}{E.~Encrenaz} \& \bibinfo{author}{L.~Fribourg}
  (\bibinfo{year}{2008}): \emph{\bibinfo{title}{Time Separation of Events\,: An
  Inverse Method}}.
\newblock In: {\sl \bibinfo{booktitle}{{P}roceedings of the {LIX} {C}olloquium
  '06}}, {\sl \bibinfo{series}{{ENTCS}}} \bibinfo{volume}{209}.
  \bibinfo{publisher}{Elsevier Science Publishers},
  \bibinfo{address}{Palaiseau, France}.
\bibitemend

\bibitemstart{howard60}
\bibinfo{author}{R.~A. Howard} (\bibinfo{year}{1960}):
  \emph{\bibinfo{title}{Dynamic Programming and Markov Processes}}.
\newblock \bibinfo{publisher}{John Wiley and Sons, Inc.}
\bibitemend

\bibitemstart{kmst59}
\bibinfo{author}{J.~Kemeny}, \bibinfo{author}{H.~Mirkil},
  \bibinfo{author}{J.~Snell} \& \bibinfo{author}{G.~Thompson}
  (\bibinfo{year}{1959}): \emph{\bibinfo{title}{Finite mathematical
  structures}}.
\newblock \bibinfo{publisher}{Prentice-Hall, Englewood Cliffs, N.J.}
\bibitemend

\bibitemstart{pbbm98}
\bibinfo{author}{G.~A. Paleologo}, \bibinfo{author}{L.~Benini},
  \bibinfo{author}{A.~Bogliolo} \& \bibinfo{author}{G.~De~Micheli}
  (\bibinfo{year}{1998}): \emph{\bibinfo{title}{Policy optimization for dynamic
  power management}}.
\newblock In: {\sl \bibinfo{booktitle}{{DAC} '98}}. \bibinfo{publisher}{ACM},
  \bibinfo{address}{New York, NY, USA}, pp. \bibinfo{pages}{182--187}.
\bibitemend

\bibitemstart{sb03}
\bibinfo{author}{L.~Stachniss} \& \bibinfo{author}{W.~Burgard}
  (\bibinfo{year}{2003}): \emph{\bibinfo{title}{The Markov Decision Problem -
  Autonomous Mobile Systems}}.
\newblock
  \urlprefix\url{http://ais.informatik.uni-freiburg.de/teaching/ss03/ams/Decis%
ionProblems.pdf}.
\newblock \bibinfo{note}{Course notes, University Freiburg, Germany}.
\bibitemend

\bibliographyend
\end{thebibliography}

\appendix


\section{Markov Decision Processes Algorithms}\label{a:mdp}

\begin{figure}[!ht]
\centering
\fbox{
\begin{minipage}{0.9\textwidth}
\noindent
{\bf ALGORITHM \mdpValueDet{}$(M, \mu)$}

\smallskip

\noindent
\begin{tabular}{l @{\ \ \ } l @{\,:\ } l}
	\emph{Input} & $M$ & Markov Decision Process $(S, A, \prob, \we)$\\
		& $\mu$ & Policy \\
	\emph{Output} & $\va$ & Value function \\ 
\end{tabular}

\medskip

{\bf SOLVE} $ \{ \va[s] = \we_{\mu[s]}(s) + \sum_{s' \in S} \prob(s, \mu[s], s') \times  \va[s'] \} _ {s \in S \setminus s_n}$

\end{minipage}
}

\caption{Algorithm for value determination for MDPs}
\label{algo:mdp:valueDet}
\end{figure}

\begin{figure}[!ht]
\centering
\fbox{
\begin{minipage}{0.9\textwidth}
\noindent
{\bf ALGORITHM \mdpPolIt{}($M$)}

\smallskip

\noindent
\begin{tabular}{l @{\ \ \ } l @{\,:\ } l}
	\emph{Input} & $M$ & Markov Decision Process $(S, A, \prob, \we)$\\
	\emph{Output} & $\mu$ & Policy optimal w.r.t. $\we$ (initially random) \\
		& $\va$ & Value function  \\ 
\end{tabular}

\medskip

{\bf REPEAT}

	\ind{1}
	\begin{affectations}
	$\va $ & $ $ \mdpValueDet{}$(M, \mu)$ \\
	$\mathit{fixpoint} $ & $ \mathit{True}$ \\
	\end{affectations}

	\ind{1}	\textbf{for each} $s \in S \setminus s_n $ \textbf{DO}

		\ind{2}
		\begin{affectations}
		  $\mathit{optimum} $ & $ \va[s]$\\
		\end{affectations}

		\ind{2} \textbf{for each} $a \in e(s)$ \textbf{DO}

			\ind{3} \textbf{IF} $\we_a(s) + \sum_{s' \in S} \prob(s, a, s') \va[s']  < \mathit{optimum}$ \textbf{THEN}

				\ind{4}
				\begin{affectations}
				 	$\mathit{optimum} $ & $ \we_a(s) + \sum_{s' \in S} \prob(s, a, s') \va[s'] $ \\
					$\mu[s] $ & $ a$\\
					$\mathit{fixpoint} $ & $ \mathit{False}$ \\
				\end{affectations}

			\ind{3} \textbf{FI}\\
		\ind{2}	\textbf{OD}

%
%
	\ind{1}	\textbf{OD}\\
{\bf UNTIL} $\mathit{fixpoint}$
\end{minipage}
}

\caption{Algorithm of policy iteration for MDPs}
\label{algo:mdp:polIt}
\end{figure}

\newpage

\section{Maximal Circuit Mean Algorithms}\label{a:maxplus}

\begin{figure}[!ht]
\centering
\fbox{
\begin{minipage}{0.9\textwidth}
\noindent
{\bf ALGORITHM $\maxPlusValueDet(M, \mu)$}

\smallskip

\noindent
\begin{tabular}{l @{\ \ \ } l @{\,:\ } l}
	\emph{Input} & $M$ & Matrix \\
		& $\mu$ & Policy \\
	\emph{Output} & $(\eta, x)$ & Eigenmode of $M^\mu$ \\
\end{tabular}

\smallskip

\begin{enumerate}
 \item Find a circuit $c$ in the graph of $M^\mu$.

 \item Set
$$ \overline{\eta} = \frac {\sum_{e \in c} w(e)} {\sum_{e \in c} 1} $$

\item Select an arbitrary state $i$ in $c$,
set $\eta_i := \overline{\eta}$,
and set $x_i$ to an arbitrary value, say $x_i := 0$.

\item Visiting all the states $j$ that have access to $i$ in backward topological order, set
\begin{eqnarray}
\eta_j & := & \overline{\eta}\\
x_j & := & \w{j}{\mu(j)} - \overline{\eta} + x_{\mu(j)} \label{maxplus:eq:21}
\end{eqnarray}

\item If there is a nonempty set $C$ of states $j$ that do not have access to $i$,
repeat the algorithm using the $C \times C$ submatrix of $M$
and the restriction of $\mu$ to $C$.
\end{enumerate}

\end{minipage}
}
\caption{Algorithm of value determination for maximal circuit mean in \maxplus{} algebras}
\label{algo:maxplus:valueDet}
\end{figure}

\begin{figure}[!ht]
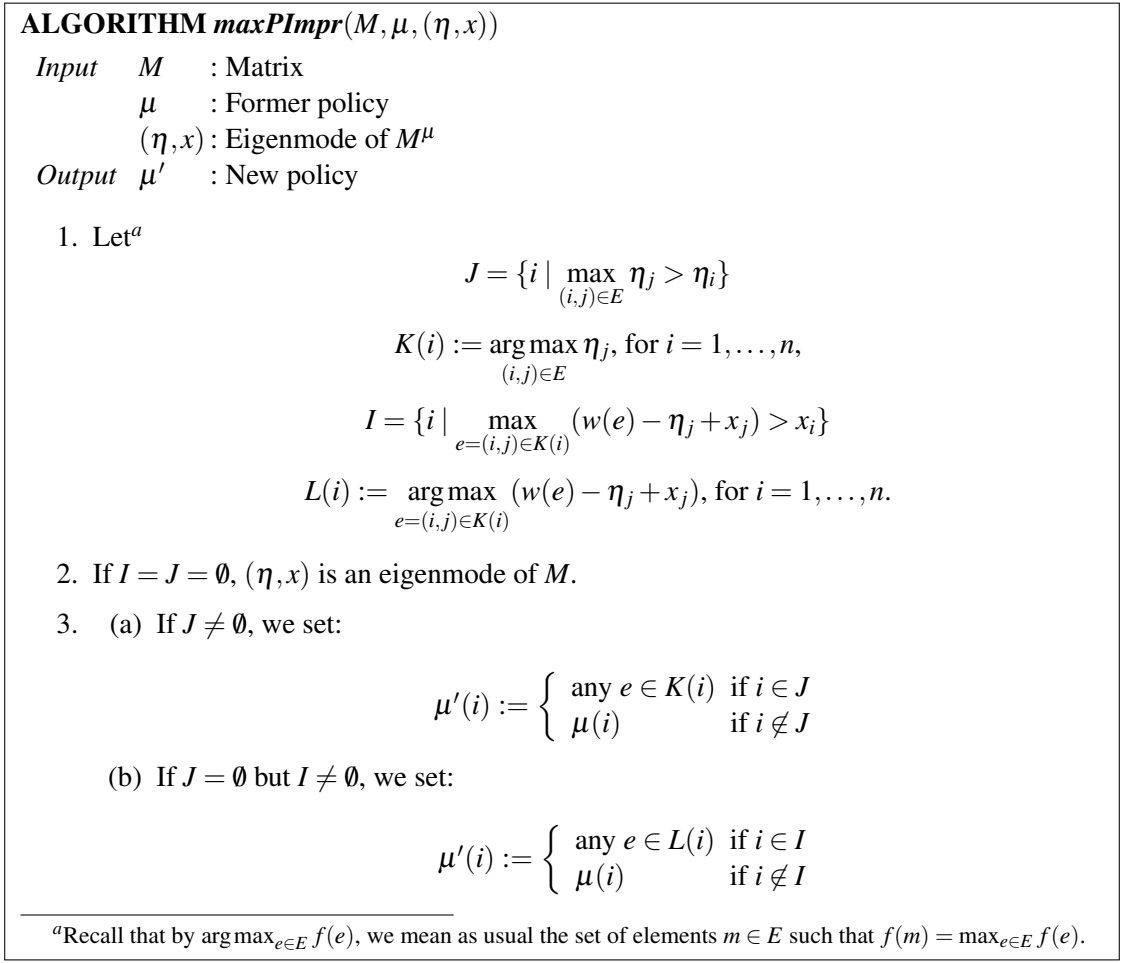

\centering
\fbox{
\begin{minipage}{0.9\textwidth}
\noindent
{\bf ALGORITHM \maxPlusPolicyImpr{}$(M, \mu, (\eta, x))$}

\smallskip

\noindent
\begin{tabular}{l @{\ \ \ } l @{\,:\ } l}
	\emph{Input} 	& $M$ & Matrix \\
			& $\mu$ & Former policy \\
			& $(\eta, x)$ & Eigenmode of $M^\mu$ \\
	\emph{Output}	& $\mu'$ & New policy \\
\end{tabular}

\smallskip

\begin{enumerate}
 \item Let\footnote{Recall
that by $\argmax_{e \in E}f(e)$, we mean as usual the set of elements
$m \in E$ such that $f(m) = \max_{e \in E}f(e)$.
}
$$ J = \{ i \mid \max_{(i,j) \in E} \eta_j > \eta_i \} $$
$$K(i) := \argmax_{(i,j) \in E} \eta_j \text{, for }i = 1, \dots, n\text{,}$$
$$ I = \{ i \mid \max_{e = (i, j) \in K(i)} (w(e) - \eta_j + x_j) > x_i \} $$
$$L(i) := \argmax_{e = (i, j) \in K(i)} (w(e) - \eta_j + x_j) \text{, for }i = 1, \dots, n\text{.}$$

\item If $I = J = \emptyset$, $(\eta, x)$ is an eigenmode of $M$.

\item
	\begin{enumerate}
	 \item If $J \neq \emptyset$, we set:\\
	$$
\begin{array}{l @{\ := \ } l}
	\mu'(i) &
	\left \{
	\begin{array}{l @{\ \ \text{if}\ } l}
	\text{any } e \in K(i) & i \in J \\
	\mu(i) & i \not\in J \\
	\end{array} \right.
\end{array}
	$$

	 \item If $J = \emptyset$ but $I \neq \emptyset$, we set:\\
	$$
\begin{array}{l @{\ := \ } l}
	\mu'(i) &
	\left \{
	\begin{array}{l @{\ \ \text{if}\ } l}
	\text{any } e \in L(i) & i \in I \\
	\mu(i) & i \not\in I \\
	\end{array} \right.
\end{array}
	$$
	\end{enumerate}
\end{enumerate}

\end{minipage}
}

\caption{Algorithm of policy improvement for maximal circuit mean in \maxplus{} algebras}
\label{algo:maxplus:policyImpr}
\end{figure}

\begin{figure}[!ht]
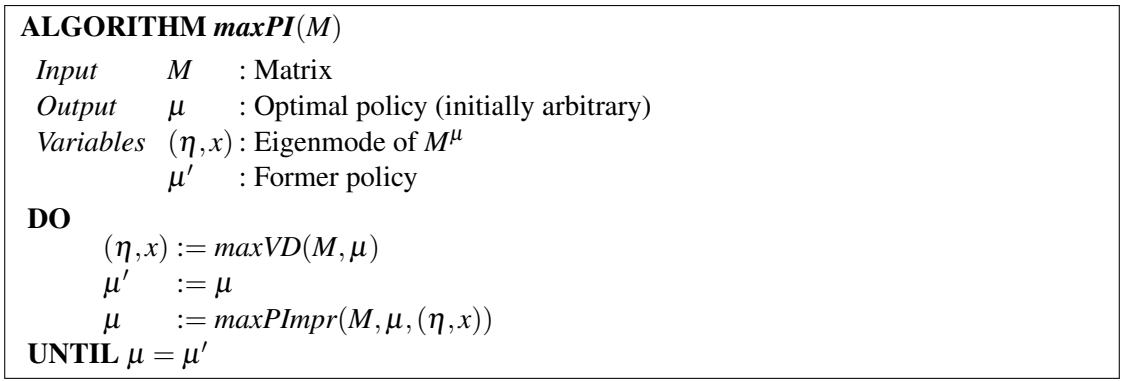

\centering
\fbox{
\begin{minipage}{0.9\textwidth}
\noindent
{\bf ALGORITHM \maxPlusPolIt$(M)$}

\smallskip

\noindent
\begin{tabular}{l @{\ \ \ } l @{\,:\ } l}
	\emph{Input} 	& $M$ & Matrix \\
	\emph{Output}	& $\mu$ & Optimal policy (initially arbitrary) \\
	\emph{Variables}& $(\eta, x)$ & Eigenmode of $M^\mu$ \\
			& $\mu'$ & Former policy\\
\end{tabular}

\smallskip

\ind{0} \textbf{DO}

	\ind{1}
	\begin{affectations}
		$(\eta, x)$ & \maxPlusValueDet$(M, \mu)$ \\
		$\mu'$ & $\mu$ \\
		$\mu$ & \maxPlusPolicyImpr$(M, \mu, (\eta, x))$ \\
	\end{affectations}

\ind{0} \textbf{UNTIL} $\mu = \mu'$

\end{minipage}
}
\caption{Algorithm of policy iteration for maximal circuit mean in \maxplus{} algebras}
\label{algo:maxplus:polIt}
\end{figure}

\end{document}